\documentclass[iop,numberedappendix]{emulateapj}
\usepackage{epsfig}
\usepackage{amsmath}
\usepackage{enumerate}
\usepackage{natbib}
\usepackage[backref,breaklinks,colorlinks,citecolor=blue]{hyperref}

\usepackage{ulem}
\usepackage{color}

\newbox\grsign \setbox\grsign=\hbox{$>$} \newdimen\grdimen \grdimen=\ht\grsign
\newbox\simlessbox \newbox\simgreatbox \newbox\simpropbox
\setbox\simgreatbox=\hbox{\raise.5ex\hbox{$>$}\llap
  {\lower.5ex\hbox{$\sim$}}}\ht1=\grdimen\dp1=0pt
\setbox\simlessbox=\hbox{\raise.5ex\hbox{$<$}\llap
  {\lower.5ex\hbox{$\sim$}}}\ht2=\grdimen\dp2=0pt
\setbox\simpropbox=\hbox{\raise.5ex\hbox{$\propto$}\llap
  {\lower.5ex\hbox{$\sim$}}}\ht2=\grdimen\dp2=0pt
\def\simgt{\mathrel{\copy\simgreatbox}}
\def\simlt{\mathrel{\copy\simlessbox}}

\newcommand{\be}{\begin{equation}}
\newcommand{\ee}{\end{equation}}
\newcommand{\sigmat}{\ensuremath{\sigma_\mathrm{T}}}
\newcommand{\xiKN}{\ensuremath{\xi_{KN}}}

\newcommand{\thd}{\theta}
\newcommand{\Zind}{Z_\star}

\newcommand{\sT}{\sigma_\mathrm{T}}
\newcommand{\epmin}{\epsilon_{\min}}
\newcommand{\epthr}{\epsilon_{\rm thr}}
\newcommand{\Nthr}{N_{\rm thr}}

\shorttitle{RMS in GRBs: Subshock photon production}
\shortauthors{Lundman \& Beloborodov}

\begin{document}

\title{Radiation mediated shocks in gamma-ray bursts: Subshock photon production}
\author{Christoffer Lundman\altaffilmark{1,2,3} and Andrei M. Beloborodov\altaffilmark{1}}
\affil{$^1$Physics Department and Columbia Astrophysics Laboratory, Columbia University, 538 West 120th Street, New York, NY 10027, USA \\
$^2$Department of Physics, KTH Royal Institute of Technology, AlbaNova, SE-106 91 Stockholm, Sweden \\
$^3$The Oskar Klein Centre for Cosmoparticle Physics, AlbaNova, SE-106 91 Stockholm, Sweden}

\label{firstpage}

\begin{abstract}
Internal shocks provide a plausible heating mechanism in the jets of gamma-ray bursts (GRBs). Shocks occurring below the jet photosphere are mediated by radiation. It was previously found that radiation mediated shocks (RMSs) inside GRB jets are inefficient photon producers, and the photons that mediate the RMS must originate from an earlier stage of the explosion. We show that this conclusion is valid only for non-magnetized jets. RMSs that propagate in moderately magnetized plasma develop a collisionless subshock which locally heats the plasma to a relativistic temperature, and the hot electrons emit copious synchrotron photons inside the RMS. We find that this mechanism is generally effective for mildly relativistic shocks and may be the main source of photons observed in GRBs. We derive a simple analytical formula for the generated  photon number per proton, $Z$, which gives $Z=10^5$-$10^6$, consistent with observations. The photons are initially injected with low energies, well below the observed GRB peak. Their number is controlled by two main factors: (1) the abundance of electron-positron pairs created in the shock, which is self-consistently calculated, and (2) the upper limit on the brightness temperature of soft radiation set by induced downscattering. The injected soft photons that survive induced downscattering gain energy in the RMS via bulk Comptonization and shape its nonthermal spectrum.
\end{abstract}

\keywords{gamma-ray burst: general -- magnetohydrodynamics (MHD) -- radiation mechanisms: thermal -- radiative transfer -- scattering -- shock waves}


\section{Introduction}
\label{sec:introduction}

Photospheric emission models are commonly invoked to explain the prompt gamma-ray burst (GRB) emission (see \citealt{BelMes:2017} for a review). Such models typically assume strong dissipation below the photosphere of the relativistic jet. The dissipated energy heats the radiation component of the jet, leading to bright, efficient prompt emission with a nonthermal spectrum consistent with observations.

An important parameter of GRBs is their emitted photon number, which regulates the position of the observed spectral peak \citep{Bel:2013}. Advection of thermal radiation from the
hot central engine provides a minimum photon number per proton, $Z=n_\gamma/n_p\sim 10^4-10^5$, which is typically insufficient to explain observations. 
This number would be conserved in adiabatic, non-dissipative, expansion, and so
dissipation in the jet is required to generate additional photons. 
In particular, photon production by
thermal bremsstrahlung and double Compton scattering can be efficient if dissipation occurs in the dense plasma at small radii where the scattering optical depth of the jet $\tau$ exceeds about $3 \times 10^4$ \citep{Bel:2013, VurLyuPir:2013}. \citet{BegPeeLyu:2017} also proposed that these processes may operate at larger radii in compressed current sheets.

However, dissipation at very large optical depths is not sufficient to explain the GRB spectrum. Most 
of the energy dissipated deep below the photosphere will be lost to neutrino emission or converted to
bulk kinetic energy as the outflow expands adiabatically, before reaching the photosphere where the radiation is released. Furthermore, any non-thermal features would be erased before radiation is released at the photosphere. Therefore, bright photospheric emission with a non-thermal spectrum requires dissipation at moderate optical depths.

In this paper, we consider a concrete dissipation process: an internal shock that propagates below the 
jet photosphere. Such shocks are mediated by photons, which shape the shock velocity profile through scatterings. Previous models of radiation-mediated shocks (RMSs) in GRB jets assumed negligible photon production by the shock itself compared with photons already available in the pre-shock plasma expanding from the hot central engine (\citealt{Lev:2012, Bel:2017} (hereafter B17), \citealt{LunBelVur:2018} (LBV17), \citealt{ItoEtAl:2017}). 
Bremsstrahlung and double Compton processes are too slow to modify the photon number in the vicinity of the shock \citep{LevBro:2008, BroMikLev:2011,Lev:2012}.

Photon production can be effective at moderate optical depths if the dissipation process generates relativistic electrons and the jet carries a magnetic field. Then synchrotron emission greatly affects the photospheric radiation spectrum (see radiative transfer simulations by \citealt{VurBel:2016}). Relativistic electrons are not generated by RMSs in non-magnetized media -- although the radiation spectrum inside the shock front extends up to the MeV band, the plasma stays at the Compton temperature of radiation, which is much smaller than 1 MeV (B17; LBV17; Ito et al. 2017). In contrast, RMSs that propagate in magnetized plasma develop a collisionless subshock (B17). The subshock dissipates a fraction of the total shock energy by impulsively heating the plasma. It heats electrons to a relativistic temperature, and they quickly emit a fraction of their energy in synchrotron photons, before cooling back to the Compton temperature. This offers a mechanism for photon production inside the RMS.

The purpose of this paper is to examine the subshock synchrotron emission
and assess under what conditions it becomes important.
We find that a key parameter controlling photon production is the dimensionless momentum of the upstream motion relative to the downstream, $\beta_r\gamma_r$. The importance of this parameter stems from the fact that it controls the multiplicity of $e^\pm$ pair creation in the RMS (LBV17; see also B17 and Ito et al. 2017).

RMSs with $\beta_r\gamma_r\simgt 1$ inevitably energize a fraction of photons to energies above the electron rest mass $m_ec^2$, 
and the RMS develops a huge $e^\pm$-to-proton ratio, $Z_\pm \equiv n_\pm/n_p \simgt 10^2$. This 
results in a relatively low thermal electron Lorentz factor behind the subshock, 
reducing the efficiency of synchrotron photon production. In contrast, mildly relativistic RMSs 
($\beta_r\gamma_r \simlt 1$) do not scatter photons to energies above $m_ec^2$.
Then inverse Compton (IC) photons from the hot subshock create $e^\pm$ pairs.
 This gives a smaller $Z_\pm$ and much more efficient production of synchrotron photons. 

The paper is organized as follows. In Section~\ref{sec:description} we provide a concise, qualitative description of the RMS and subshock structure. The subshock synchrotron source is considered in detail in Section~\ref{sec:synchrotron}, and in Section~\ref{sec:absorption} we examine absorption of the synchrotron photons in the cooler plasma outside the subshock region. In Section~\ref{sec:surviving_photons}, we estimate the number of surviving synchrotron photons that populate the RMS. In Section~6 we apply the results to GRB shocks. Our conclusions are summarized in Section~\ref{sec:discussion}.

\section{Qualitative description of the RMS}
\label{sec:description}

Consider a shock that propagates in an optically thick magnetized plasma. Such shocks will develop a collisionless subshock (B17). The presence of the subshock can be qualitatively understood as follows. The downstream energy densities of radiation and magnetic field are determined by the shock jump conditions. The hot downstream photons diffuse into the upstream, attempting to decelerate the incoming upstream as they scatter. However, photons carry only a fraction of the total downstream energy density (the energy is partially stored in the compressed magnetic field), and therefore they can only partially decelerate the incoming upstream flow. A collisionless subshock must develop close to the downstream side of the RMS in order to dissipate the remaining incoming kinetic energy.

\begin{figure}
\includegraphics[width=\linewidth]{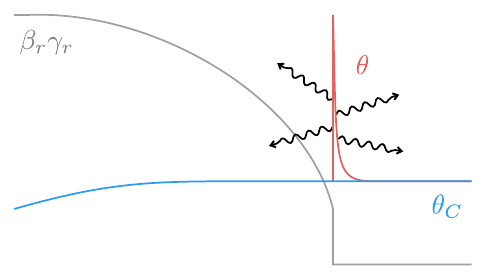}
\caption{A schematic view of the shock structure. The flow moves from left to right. The gray line shows the (dimensionless) momentum $\beta\gamma$ relative to the downstream frame. It equals $\beta_r\gamma_r$ far upstream and decreases inside the RMS. The subshock is located in the immediate RMS downstream and indicated by the discontinous jump in $\beta\gamma$. The red line shows the electron (or pair) temperature $\theta = kT / m_e c^2$. The hot electrons cool by IC
scatterings and synchrotron emission, and can act as a photon source for the RMS. The electrons cool until they reach the Compton temperature $\theta_C$ of the local radiation field; $\theta_C$ is shown by the blue line. The width of the cooling region indicated by the red line is much smaller than the RMS width.}
\label{fig:schematic_structure}
\end{figure}

A schematic view of the flow profile is shown in Figure~\ref{fig:schematic_structure}. The RMS has a width of at least a few scattering mean free paths. The subshock width is microscopic in comparison. It is comparable to the proton Larmor radius, which is many order of magnitudes smaller than the photon scattering mean free path. The plasma particles are impulsively heated by the subshock. 
Magnetic fields frozen in expanding flows are expected to become transverse to the flow velocity 
and parallel to internal shocks. Then Fermi acceleration of particles is inefficient and they form 
quasi-Maxwellian
distribution. The electron component typically obtains a fraction $0.3-0.5$ of the dissipated (subshock) energy \citep{SirSpi:2011}.\footnote{Collisionless shocks in electron-ion plasma loaded with copious $e^\pm$ pairs have not been studied  in detail yet; we will assume that in this case the $e^\pm$ receive a similar large fraction of the shock energy.} Behind the subshock the heated $e^\pm$ will quickly radiate their energy due to inverse Compton and synchrotron emission. 
 
We will refer to the high-temperature region of the flow behind the subshock as the ``cooling region'' or the ``source.'' Synchrotron photons supplied by the source can participate in mediating the RMS and increase their energies via bulk Comptonization by the plasma with the velocity profile shown in Figure~\ref{fig:schematic_structure}.

The synchrotron spectrum emitted
by the subshock is essentially that expected from a fast-cooling thermal $e^\pm$ population. A few important characteristic frequencies can be identified. The frequency

\be
\label{eq:nnu}
\nu_0 \approx \gamma_0^2 \nu_B, \qquad \nu_B \equiv \frac{e B}{2\pi m_e c},
\ee

\noindent corresponds to the (thermal) electron Lorentz factor $\gamma_0$ just behind the subshock. Here $e$ is the electron charge, $B$ is the magnetic field strength, $m_e$ is the electron rest mass and $c$ is the speed of light. The synchrotron emission will become self-absorbed at some frequency $\nu_{bb}$. We refer to the partially self-absorbed synchrotron spectrum emitted from the cooling region as the ``source spectrum.''

\begin{figure}
\includegraphics[width=\linewidth]{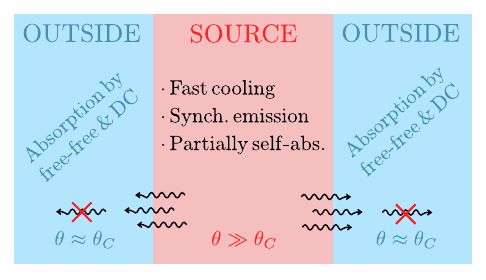}
\caption{A schematic view of the different regions relevant to subshock photon production. The subshock cooling region acts as a source of (low-energy) synchrotron photons (pink region; it was indicated by the red curve in Figure~\ref{fig:schematic_structure}). The photon source is self-absorbed at low frequencies. In order to populate the RMS, photons escaping the source must also survive free-free and double Compton absorption in the much more extended (blue) region of the RMS outside the synchrotron source.}
\label{fig:regions}
\end{figure}

\medskip

Figure~\ref{fig:regions} highlights the most relevant properties of the source region and its surroundings. The source spectrum is emitted into the surrounding plasma (blue region in Figure~\ref{fig:regions}), which is locked to the local Compton temperature of the RMS radiation. The synchrotron photons have low energies, and only a fraction of them may be able to avoid free-free
absorption in the blue region. The synchrotron radiation has a high brightness temperature, and therefore experiences induced downscattering (discussed in more detail in Section~\ref{ssec:induced_scattering}). The downscattered photons are inevitably absorbed by the plasma. 
The competition between absorption and bulk Comptonization in the RMS defines an effective absorption frequency $\nu_\star$, above which photons will survive and populate the RMS.

\begin{figure}
\includegraphics[width=\linewidth]{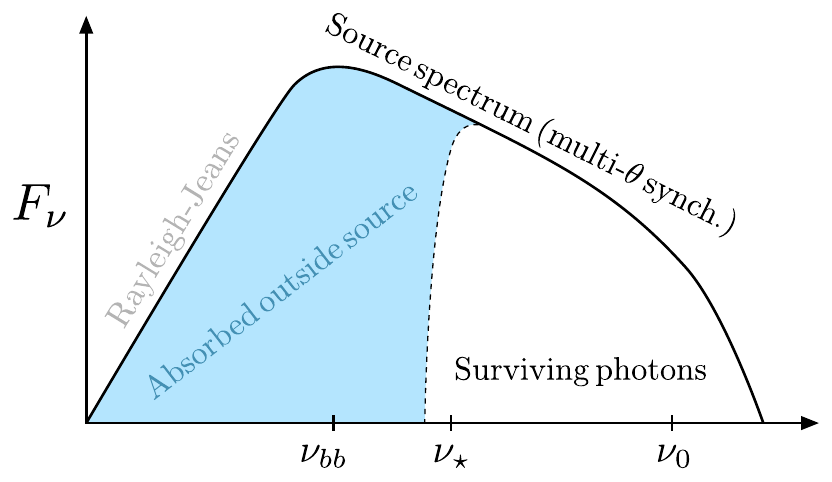}
\caption{An illustration of the synchrotron spectrum emerging from the source (pink region in Figure~\ref{fig:regions}) and then propagating through the RMS plasma.
The spectrum is self-absorbed inside
the source at $\nu < \nu_{\rm bb}$; this shapes the Raleigh-Jeans slope at low frequencies. 
Radiation with $\nu < \nu_\star$ is absorbed by the plasma in the RMS outside the 
synchrotron source, as a result of induced downscattering and free-free absorption. 
The figure assumes $\nu_{bb} < \nu_\star < \nu_0$; the regime $ \nu_\star > \nu_0$  
is also possible, as will be discussed below in the paper.}
\label{fig:schematic_spectrum}
\end{figure}

The number of photons that populate the RMS depends 
on the ordering of $\nu_{bb}$, $\nu_\star$ and $\nu_0$. A schematic illustration of the 
generated synchrotron
spectrum with the frequency ordering $\nu_{bb} < \nu_\star < \nu_0$ is shown in Figure~\ref{fig:schematic_spectrum}.

\section{The synchrotron source}
\label{sec:synchrotron}

\subsection{Fast cooling of subshock plasma}

The cooling time for an electron with Lorentz factor $\gamma \gg 1$ immersed in photons and magnetic fields with energy densities $u_\gamma$ and $u_B$ respectively is

\be
t_c = \frac{3}{4} \frac{m_e c}{\gamma \sigmat (\xiKN u_\gamma + u_B)},
\ee

\noindent where $\sigmat$ is the Thomson cross-section and $\xiKN \leq 1$ is a correction factor due to possible Klein-Nishina effects; $\xi_{KN}\approx 1$ for GRB jets. It is convenient to define the (dimensionless) enthalpies associated with photons and magnetic fields

\be
w \equiv \frac{u_\gamma + p_\gamma}{\rho c^2} = \frac{4}{3} \frac{u_\gamma}{\rho c^2},
\ee

\noindent and

\be
\sigma \equiv \frac{B^2}{4 \pi \rho c^2} = 2 \frac{u_B}{\rho c^2},
\ee

\noindent where 
$\rho\approx n_p m_p$ is the plasma rest-mass density. It is useful to compare the electron cooling time with the mean free path time of photons to Thomson scattering, $t_{sc}$,

\be
\frac{t_c}{t_{sc}} = \frac{m_e}{m_p} \frac{Z_\pm g}{\gamma w \xiKN},
\label{eq:times_ratio}
\ee

\noindent where the scattering time is $t_{sc} = (Z_\pm n_p \sigmat c)^{-1}$, $Z_\pm \equiv n_\pm/n_p$ is the $e^\pm$-to-proton ratio and

\be
g \equiv \frac{1}{1 + u_B / \xiKN u_\gamma}
\label{eq:g}
\ee

\noindent is a numerical factor, which is close to unity if $u_B < u_\gamma$. The values of $u_\gamma$ and $u_B$ at the location of the subshock are comparable to their values in the far downstream, which are given by the RMS jump conditions.

Using $g \sim 1$ and $\xi_{KN} \approx 1$ in Equation~(\ref{eq:times_ratio}) one can see that $t_c \ll t_{sc}$ for $Z_\pm\ll \gamma w m_p/m_e$, which is generally satisfied. A conservative upper limit on $Z_\pm$ generated by a shock is set by its energy budget: $Z_\pm m_e = (3/4) w m_p$ would correspond to all dissipated energy being converted to pair rest mass; the actual $Z_\pm$ is well below this upper limit, and hence $Z_\pm \ll \gamma w m_p/m_e$.

The fact that $t_c\ll t_{sc}$ implies that the photon field does not change much across the electron cooling length behind the subshock. Therefore, one can view the synchrotron source as a thin layer of high-energy electrons/positrons placed in a uniform external radiation field. The width of the synchrotron source is smaller than the RMS width by the factor $\sim t_c/t_{sc}$ given in Equation~(\ref{eq:times_ratio}).

\subsection{Pair loading reduces $\gamma_0$}
\label{ssec:pair_loading}

Next, we define $f$ as the fraction of the shock energy that is dissipated in the subshock; $0 < f < 1$ is a measure of the subshock strength. This fraction is itself not a free parameter; it depends on the shock parameters $w$, $\sigma$ (measured in the downstream) and $\beta_r\gamma_r$. This dependence has not been calculated, however $f$ is known to grow
with increasing magnetization and approach unity for shocks with downstream $\sigma > w$ (B17).

The thermal energies of subshock-heated $e^\pm$ particles $\sim (\gamma_0 - 1) m_e c^2$ convert to radiation as the particles cool, producing energy density $f u_\gamma$, which may be written as $f u_\gamma \approx (\gamma_0-1)Z_\pm n_p m_ec^2$. As long as $\gamma_0\gg 1$ this gives

\be
\gamma_0 Z_\pm \approx \frac{3}{4}\frac{m_p}{m_e} f w \approx 1400 f w.
\label{eq:gamma_0_Z_pm}
\ee

\noindent Here the downstream $w = (4/3)u_\gamma/n_pm_pc^2$ may be estimated using the RMS jump conditions.\footnote{A more accurate estimate of $w$ in the immediate downstream should take into account that about half of the energy dissipated in the subshock is stored in protons. The proton energy converts to radiation with a delay, in the far downstream, and thus does not contribute to $u_\gamma$ in the immediate downstream (see B17).} Equation~(\ref{eq:gamma_0_Z_pm}) shows that a high $Z_\pm$ implies a low $\gamma_0$. This is simply because the shock energy is shared among a larger population of $e^\pm$, and so the energy per particle is reduced. 

The subshock synchrotron emission can be efficient when $\gamma_0$ is sufficiently large. Since $\gamma_0 \propto Z_\pm^{-1}$, strong pair loading threatens efficient synchrotron emission.

\subsection{Two possible sources of pairs}

Two mechanisms can generate MeV gamma-rays that create $e^\pm$ pairs in an RMS (B17): (1) bulk Comptonization of photons by the converging velocity profile of the RMS and (2) IC emission from the collisionless subshock. Below we discuss the two mechanisms in turn. 

Bulk Comptonization results from the upstream motion with velocity $\beta_r$ relative to the downstream, and its efficiency depends on $\beta_r$. Bulk Comptonization in non-relativistic shocks, $\beta_r \ll 1$, is slow and incapable of upscattering photons to MeV energies, because their energy gain per scattering becomes unable to compete with increasing loss to electron recoil. As a result the Comptonized spectrum extends only to

\be
\epsilon_{\max} \sim \beta_r^2, \qquad \epsilon \equiv \frac{E}{m_ec^2},
\ee

\noindent 
where $E$ is the photon energy.

Relativistic RMSs, $\beta_r\gamma_r \simgt 1$, are qualitatively different (LBV17); then bulk Comptonization successfully generates MeV gamma-rays, which leads to copious pair creation. The RMS becomes heavily loaded with $e^\pm$, $Z_\pm \gg 1$, regardless of the presence or absence of the subshock, and so regardless of magnetization. B17 estimated that bulk Comptonization in internal shocks in GRB jets generates a characteristic $Z_\pm \sim 10^2$, and LBV17 explicitly demonstrated the efficiency of pair creation by tracking photon-photon collisions $\gamma\gamma\rightarrow e^+e^-$ in first-principle simulations of the RMS structure. In particular, LBV17 found $Z_\pm\approx 220$ in an RMS with $\beta_r\gamma_r = 3$ and an assumed
photon-to-proton ratio of $n_\gamma/n_p = 2 \times 10^5$.

Next, we describe pair creation by the IC photons emitted by the collisionless subshock. This mechanism operates even in non-relativistic RMSs, as long as their magnetization is sufficient to sustain a strong subshock. Pairs are produced if the subshock thermal $\gamma_0$ is high enough to generate MeV gamma-rays through IC scattering. This situation is expected when pair loading through bulk Comptonization is inefficient, i.e. when $\beta_r\gamma_r\simlt 1$.

The width of the cooling region behind the subshock is much smaller than the photon mean free path to both scattering and $\gamma\gamma$ absorption (B17, LBV17). 
When viewed from the downstream plasma, the subshock is moving away at a speed $\beta_d$.
IC photons emitted at angles $\cos\theta > \beta_d$ with respect to the shock normal will
propagate ahead of the subshock and create pairs there. These IC photons are emitted within a solid angle $2\pi(1-\beta_d)$, which is 0.3-0.5
of the full $4\pi$. Therefore, one can estimate that a fraction 
$\psi=0.3$-0.5 of IC photons overtake the subshock and can convert to pairs there. These pairs
are injected into the plasma upstream of the subshock, and then advected through the subshock and heated to $\gamma_0$.

The upstream pair creation quickly grows until saturation is reached. The saturation occurs because the growing $Z_\pm$ reduces $\gamma_0$, which makes the subshock IC photons less energetic and thus less capable of producing $e^\pm$. In the resulting self-regulated state each downstream pair produces one pair upstream of the subshock. B17 estimated the self-regulated value of $\gamma_0 \sim 20$.

The exact $\gamma_0$
depends on the photon spectrum in the immediate RMS downstream, which is in general highly non-thermal. As shown by LBV17, the spectrum inside a strong RMS ranges typically from flat in $\nu F_\nu$ (constant energy per decade in photon energy) to flat in $F_\nu$ (constant photon number per decade in photon energy). The spectrum extends from a lower energy $\epsilon_{min}$ to an upper energy $\epsilon_{max}$. For mildly relativistic shocks, Klein-Nishina effects reduce the effectiveness of photon energy gain, causing the effective upper energy to be $\epsilon_{max} \sim 0.1$.

The calculation of the self-regulated $\gamma_0$ can be set up as follows. Consider 
an electron injected at the subshock with a Lorentz factor $\gamma_0$ and immersed in radiation 
with intensity $\propto\epsilon_t^{-\alpha}$ --- a target radiation spectrum extending from 
$\epsilon_t\sim\epsilon_{\min}$ to some $\epsilon_{\max}>\gamma_0^{-1}$ 
($\epsilon_{\max}$ will not enter the result). The electron experiences IC and synchrotron 
losses and its $\gamma$ quickly decreases below $\gamma_0$ behind the subshock. 
IC scattering generates photons with energies $\epsilon\approx \gamma^2\epsilon_t$,
but not exceeding the electron energy $\gamma-1$.
In the simplest approximation, this cutoff can be treated by assuming that all
IC photons with $\epsilon\simlt\gamma/2$ have been scattered with a constant Thomson 
cross section, neglecting Klein-Nishina suppression, and the IC spectrum is completely suppressed
at $\epsilon>\gamma/2$. 

Then the IC spectrum produced by an electron cooling from $\gamma$ to $\gamma-\delta\gamma$ is 
given by 
\be
     \delta \frac{dN}{d\epsilon}=\delta K\,\epsilon^{-\alpha-1}, 
  \qquad \gamma^2\epmin<\epsilon<\frac{\gamma}{2}.
\ee
Energy contained in this spectrum equals $g\,\delta\gamma$, where $g\leq 1$ is the IC fraction 
in the total (IC+synchrotron) losses; this condition gives 
\be
  \delta K=\frac{(1-\alpha)\,g\,\delta\gamma}{(\gamma/2)^{1-\alpha}-(\gamma^2\epmin)^{1-\alpha}}.
\ee

We are interested in IC photons with $\epsilon>\epthr\approx 2$, which are capable of 
converting to $e^\pm$ pairs ahead of the subshock. The number of such photons (per  
cooling electron/positron) is 
\be
  \delta \Nthr =\frac{\delta K}{\alpha}\left[\epthr^{-\alpha}-(\gamma/2)^{-\alpha}\right].
\ee 
Integrating over all $\gamma>2\epthr$ contributing to IC emission at $\epsilon>\epthr$, 
we obtain
\be
   \Nthr=g\,\frac{(1-\alpha)}{\alpha}\int_{2\epthr}^{\gamma_0} 
        \frac{\left[\epthr^{-\alpha}-(\gamma/2)^{-\alpha}\right]}
               {[(\gamma/2)^{1-\alpha}-(\gamma^2\epmin)^{1-\alpha}]}\,d\gamma.
\ee
The self-regulated $\gamma_0$ can now be determined from the condition $\psi\Nthr=1$, where 
$\psi=0.3-0.5$ is the fraction of IC photons overtaking the shock (B17).

For instance, for $\alpha=1/2$, and in the limit of $\epmin\rightarrow 0$, we find
\be
    \Nthr=2g\left(\gamma_0^{1/2}-2-\ln\frac{\gamma_0}{4}\right)=\psi^{-1}\sim 3,
\ee
where $g\approx 1$, as long as synchrotron losses are subdominant compared to IC losses.
This gives $\gamma_0\sim 30$. Similar values of $\gamma_0=20-40$ are obtained for 
other slopes $\alpha$ in the range of main interest here $0<\alpha<1$. Therefore, we estimate
for mildly relativistic RMS,
\be
  \gamma_0 \approx 20 - 40, \qquad \beta_r\gamma_r\simlt 1,
\label{eq:SS_gamma_0}
\ee
and in our estimates below $\gamma_0$  will be normalized to 30. 
The corresponding self-consistent pair loading factor $Z_\pm$ follows from 
Equation~(\ref{eq:gamma_0_Z_pm}): $Z_\pm \approx 46 \, f w (\gamma_0/30)^{-1}$.

\subsection{Synchrotron self-absorption}

Synchrotron emission from the pairs cooling behind the subshock will be self-absorbed at low frequencies $\nu < \nu_{bb}$, which we estimate below.

The energy density of synchrotron radiation $u_s$ peaks near the frequency $\nu_0 = \gamma_0^2 \nu_B$ and can be estimated as

\be
u_s(\nu_0) \approx f u_\gamma \left(1 - g\right) = \frac{f \, g}{\xiKN}\, u_B,
\label{eq:SSA_1}
\ee

\noindent where $1-g$ is the fraction of the electron energy that converts to synchrotron radiation;
the IC fraction $g$ is given in Equation~(\ref{eq:g}) and here
evaluated at $\gamma = \gamma_0$. The fast-cooling synchrotron spectrum scales as $\nu I_\nu \propto \nu^{1/2}$, and hence the energy density of photons emitted around a given frequency $\nu < \nu_0$ is

\be
u_s(\nu) = \nu\,\frac{\mathrm{d}u_s}{\mathrm{d}\nu} \approx (\nu/\nu_0)^{1/2} u_s(\nu_0).
\label{eq:SSA_2}
\ee

\noindent The emission becomes self-absorbed at the frequency where

\be
u_s(\nu) \approx u_{bb}(\nu).
\label{eq:SSA_3}
\ee

\noindent Here $u_{bb}(\nu) = \nu (\mathrm{d}u_{bb}/\mathrm{d}\nu)$ is the energy density of blackbody radiation of temperature $\theta_\nu\approx \gamma_\nu/3$ that corresponds to the energy of electrons emitting at $\nu$, $\gamma_\nu\approx (\nu/\nu_B)^{1/2}$. The synchrotron photons have energies $h\nu \ll kT_\nu = \theta_\nu m_e c^2$ and one can use the Rayleigh-Jeans expression,

\begin{equation}
u_{bb}(\nu)\approx \frac{8\pi m_e\nu^3\theta_\nu}{c} \approx \frac{8\pi m_e\nu^{7/2}}{3c\nu_B^{1/2}}.
\label{eq:SSA_4}
\end{equation}

\noindent Solving Equations~(\ref{eq:SSA_1}), (\ref{eq:SSA_2}), (\ref{eq:SSA_3}) and (\ref{eq:SSA_4}) for $\nu$, we find

\be
\nu_{bb} \approx \left(\frac{3}{8\pi}\,\frac{f\,g}{\xi_{KN}}\,\frac{c\,u_B}{m_e\gamma_0}\right)^{1/3}.
\ee

\noindent 
The value of $\nu_{bb}$ depends on $u_B=B^2/8\pi$
(or equivalently on $\rho$, if $\sigma$ is kept constant). This is in contrast to the shock jump conditions, which can be cast in terms of dimensionless quantities $\sigma$, $w$ and $\beta_r\gamma_r$. 

Note that the estimate for $\nu_{bb}$ was obtained assuming $\nu_{bb}<\nu_0$, 
and the result should be compared with $\nu_0$ (given in Equation~(\ref{eq:nnu})).
The ratio $\nu_{bb}/\nu_0$ scales as $B^{-1/3}$ and we can
define $B_{bb}$ as the magnetic field for which $\nu_{bb} = \nu_0$, 

\be
\frac{\nu_{bb}}{\nu_0} = \left(\frac{B_{bb}}{B}\right)^{1/3}.
\label{eq:nu_bb}
\ee

\noindent Solving for $B_{bb}$ gives

\be
\label{eq:Bbb}
B_{bb} = \frac{3}{64\pi^2}\,\frac{f\,g}{\xi_{KN}}\,\frac{m_e^2c^4}{e^3\,\gamma_0^7} \approx 3 \times 10^5 \,\frac{f\,g}{\xi_{KN}} \left(\frac{\gamma_0}{30}\right)^{-7} \, \mathrm{G}.
\ee

\noindent The subshock synchrotron source is suppressed if $B < B_{bb}$. Note the strong dependence of $B_{bb}$ on $\gamma_0$.

\section{Absorption outside the synchrotron source}
\label{sec:absorption}

The partially self-absorbed synchrotron photons will leave the cooling region behind the subshock and propagate into the larger RMS structure. The synchrotron photons have low energies, and are therefore susceptible to absorption in the plasma that surrounds the synchrotron source. The photons will take part in mediating the shock structure if they can double their energies via bulk Comptonization before they are absorbed by the plasma. This condition will determine the number of synchrotron photons injected in the RMS.

\subsection{Free-free absorption}

The typical absorption time for a photon of frequency $\nu$ is 
\be
  t_{abs}(\nu) = \frac{1}{\alpha_\nu c}, 
\ee
where $\alpha_\nu$ is the absorption coefficient of the plasma that surrounds the subshock. This 
time should be compared with the residence time of photons in the RMS, $t_{RMS}$, 
the timescale for gaining energy via bulk Comptonization. 
The average photon energy gain per scattering in a non-relativistic RMS is
$\Delta\epsilon/\epsilon \sim \beta_r^2$ (e.g. \citealt{BlaPay:1981b}; LBV17).
The same estimate is approximately valid for highly relativistic shocks, which have 
$\Delta\epsilon/\epsilon \sim 1$. The average number of scatterings experienced by a photon in 
the RMS, $N_{sc}=t_{RMS}/t_{sc}$, corresponds to Compton parameter
$y=N_{sc} \Delta\epsilon/\epsilon\sim 1$. Therefore, 
$t_{RMS}$ is related to the photon mean free path time $t_{sc}=(Z_\pm n_p\sT c)^{-1}$ by
\be
t_{RMS} \approx \frac{t_{sc}}{\beta_r^2}.
\ee
The effective absorption optical depth is defined by

\be
\tau_{abs}(\nu) \equiv \frac{t_{RMS}}{t_{abs}(\nu)} = \frac{\alpha_\nu}{n_\pm \sigmat \beta_r^2}.
\label{eq:tau_abs}
\ee

\noindent The subshock photon source is attenuated by a factor $\exp(-\tau_{abs}(\nu))$ before engaging in efficient bulk Comptonization, and we can find $\nu_{abs}$ from $\tau_{abs}(\nu_{abs}) \approx 1$.

The absorption coefficient $\alpha_\nu$ is sensitive to the plasma temperature in the RMS, $\thd$, which is locked to the Compton temperature of the radiation field, $\thd \approx \theta_C \ll 1$. 
In Appendix~\ref{app:BrDC} we discuss two relevant processes: bremsstrahlung (free-free) and double Compton scattering. The corresponding absorption coefficients are obtained from Kirchhoff's law, $\alpha_\nu = j_\nu/B_\nu$, where $j_\nu = (\hbar/2) \mathrm{d}\dot{n}_\gamma/\mathrm{d}\ln\epsilon$ is the emissivity given in Appendix~\ref{app:BrDC} and $B_\nu = (2h\nu^3/c^2)[\exp(h\nu/kT)-1]^{-1}$ is the Planck function. 

The free-free process is the fastest, so only this process is kept in the estimates below.
Its absorption coefficient in a pair-dominated plasma ($Z_\pm \gg 1$) with temperature $\thd \ll 1$ is given by

\be
\alpha_\nu \approx \frac{c^2 r_e \sigma_{\rm T} \Lambda\, n_\pm^2}{2\pi^{3/2}\,\thd^{3/2}\nu^2},
\label{eq:alpha_nu}
\ee

\noindent where $r_e=e^2/m_ec^2=2.82\times 10^{-13}$~cm is the classical electron radius,
and $\Lambda = \ln(4\thd/\epsilon) \approx 10$. From $\alpha_\nu \propto \nu^{-2}$ 
one can see that $\tau_{abs}(\nu) \propto \nu^{-2}$, and so

\be
\tau_{abs}(\nu) = \left(\frac{\nu_{abs}}{\nu}\right)^2.
\label{eq:tau_abs_2}
\ee

\noindent We find $\nu_{abs}$ by setting $\tau_{abs} = 1$ in Equation~(\ref{eq:alpha_nu}),

\be
\nu_{abs} \approx \left(\frac{\Lambda\, c^2 r_e n_\pm}{2\pi^{3/2} \thd^{3/2} \beta_r^2}\right)^{1/2}.
\ee

\noindent The absorption frequency should be compared with $\nu_0 = \gamma_0^2 \nu_B$,

\be
\label{eq:nu_abs}
\frac{\nu_{abs}}{\nu_0} \approx \left( \frac{\Lambda\,Z_\pm m_e} {2\sqrt{\pi}\, m_p\,\beta_r^2\, \thd^{3/2} \gamma_0^4\,\sigma}\right)^{1/2}.
\ee

\noindent 
If $\nu_{abs} > \nu_0$ then an exponentially small fraction of synchrotron photons avoid absorption and gain significant energies in the RMS.

Since $Z_\pm$ and $\gamma_0$ are related by Equation~(\ref{eq:gamma_0_Z_pm}), the ratio $\nu_{abs}/\nu_0$ may be rewritten in two equivalent forms,

\begin{eqnarray}
\nonumber
  \frac{\nu_{abs}}{\nu_0} &\approx& \left( \frac{ 3\Lambda}{8\sqrt{\pi}}\right)^{1/2} \frac{(fw/\sigma)^{1/2}}{\beta_r\, \thd^{3/4} \gamma_0^{5/2}} \\
  & \approx & 0.0093\,\frac{f^{1/2} w^{1/2}}{\sigma^{1/2}\beta_r\,\theta_{-2}^{3/4}} \left(\frac{\gamma_0}{30}\right)^{-5/2}. 
  \label{eq:abs}
\end{eqnarray}
\begin{eqnarray}
\nonumber
\frac{\nu_{abs}}{\nu_0} & \approx & \left(\frac{128\Lambda}{81\sqrt{\pi}}\right)^{1/2} \frac{(Z_\pm m_e/m_p)^{5/2}}{f^2 w^2 \sigma^{1/2}\beta_r\, \thd^{3/4}} \\ 
   & \approx & 0.066\,\frac{(Z_\pm/100)^{5/2}}{f^2w^2 \sigma^{1/2}\beta_r\,\theta_{-2}^{3/4}}. 
\label{eq:absZ}
\end{eqnarray}
Here we normalized the plasma (Compton) temperature $\thd\approx\theta_{C}$ to $10^{-2}$.
The Compton temperature is related to the average energy of photons heated by the shock,
\be 
  \epsilon_d\, m_ec^2 = \frac{3}{4}\frac{w\, m_p c^2}{Z}, 
 \qquad Z \equiv \frac{n_\gamma}{n_p},
\label{eq:eps_d}
\ee
where $Z$ is the photon number per proton. Far downstream, where radiation relaxes to a Wien spectrum, $\theta_C=\epsilon_d/3\approx 4.6\times 10^{-3}\,w\,Z_{5}^{-1}$. Inside the RMS the radiation spectrum has a nonthermal tail and hence a somewhat higher Compton temperature 
$\theta_C>\epsilon_d/3$; its exact value depends on the shape of the radiation spectrum. 
Therefore, we roughly estimate $\theta\sim\epsilon_d$.

\subsection{Induced downscattering}
\label{ssec:induced_scattering}

The partially self-absorbed synchrotron source emits radiation with large photon occupation numbers $N \gg 1$ and brightness temperatures $\theta_b = \epsilon N \gg 1$. Induced scattering therefore becomes important.

The brightness temperature of synchrotron radiation emitted by the subshock is given by
\be
\label{eq:thb}
  \theta_b\approx\theta_0\times \left\{\begin{array}{ll}
        1, & \;\; \nu < \nu_{bb} \\
    (\nu/\nu_{bb})^{-5/2}, & \; \nu_{bb}<\nu<\nu_0
                                                           \end{array}
                                                    \right.
\ee
Here $\theta_0$ is the maximum brightness temperature, reached at the self-absorption
frequency $\nu_{bb}$ for emission from the hottest $e^\pm$ behind 
the subshock. It equals the temperature of the hottest electrons,
\be
\label{eq:th0}
   \theta_0\approx \frac{\gamma_0}{3}. 
\ee
In Equation~(\ref{eq:thb}) we used the slope of the fast cooling synchrotron 
spectrum, $I_\nu\propto \nu^{-1/2}$ at $\nu_{bb}<\nu<\nu_0$, and the fact that brightness 
temperature scales as $I_\nu/\nu^2$.

As the synchrotron photons exit the source, they enter plasma of a significant optical
depth and a relatively low (Compton) temperature $\thd \sim \epsilon_d \ll \theta_b$.
Then induced scattering tends to lower the photon energies \citep{ZelLev:1969},
which leads to their more efficient absorption by the free-free process.
Let $\nu_{ind}$ be the characteristic frequency below which induced downscattering 
is efficient. 

As radiation propagates through a cool plasma of Thomson optical depth $\tau<1$,
induced scattering has a strong impact at frequencies where $\theta_b(\nu)\,\tau\simgt 1$. 
For a plasma  with $\tau>1$, induced scattering is expected to be important when 
$\theta_b(\nu)\,\tau^2\simgt1$. The additional factor of $\tau$ appears in this condition 
because the time it takes the photons to diffuse through the optically thick plasma is 
increased by the factor of $\tau$. In an RMS, induced downscattering will compete 
against bulk Comptonization.

The condition $\theta_b(\nu)\,\tau^2\sim 1$ introduces a characteristic frequency $\nu_{ind}$
below which induced downscattering is efficient,
\be
\label{eq:down}
   \theta_b(\nu_{ind})\sim \tau^{-2}\sim\beta_r^2.
\ee
Here $\tau\sim \beta_r^{-1}$ is the optical depth of the RMS. 
Using Equation~(\ref{eq:thb}) we find
\be
\label{eq:nuind}
 \frac{\nu_{ind}}{\nu_{bb}}\sim \left(\frac{\beta_r^2}{\theta_0}\right)^{-2/5}
 \approx 2.5\, \left(\frac{\gamma_0}{30}\right)^{2/5} 
    \beta_r^{-4/5}.
\ee

\noindent This estimate neglects factors that may reduce downscattering efficiency (reduce $\nu_{ind}$).
In particular, subtle effects may occur near the synchrotron source in the collisionless 
subshock. It emits significant low-frequency radiation into the upstream,
across the plasma velocity jump. 
The propagating radiation is directed away from the source, 
however induced scattering quickly, after passing $\delta\tau\sim\theta_0^{-1}\sim 0.1$, 
amplifies the seed (spontaneously scattered) radiation in the opposite direction 
\citep{CopBlaRee:1993}. Induced scattering not only steals energy from photons but also helps 
them to ``reflect'', so that the upward and downward fluxes of low-frequency radiation
become approximately equal in the local plasma frame. As a result, a significant fraction of 
synchrotron photons should quickly turn around and go back into the downstream. 
Their trip is accompanied by both energy loss (the upstream plasma is heated by induced 
scattering) and energy gain (the Fermi process operates in crossing the subshock ---
bulk Comptonization). Our conservative estimate of $\nu_{ind}$ in Equation~(\ref{eq:nuind}) assumes that 
the energy gain does not save the low-frequency photons from downscattering.

If induced downscattering is efficient at $\nu_0$ ($\nu_{ind}>\nu_0$) while free-free
absorption is inefficient ($\nu_{abs}<\nu_0$) then an approximate
balance will be reached between the supply of photons $\nu\sim \nu_0$ and their downscattering rate.
As a result, the brightness temperature saturates at $\theta_b^{\max}\sim\tau^{-2}$,
\be
\label{eq:cond}
  \theta_b(\nu_0)\sim\theta_b^{\max}\sim \beta_r^2  \qquad {\rm if~~} \nu_{abs}<\nu_0<\nu_{ind}.
\ee

\section{The number of surviving synchrotron photons}
\label{sec:surviving_photons}

\subsection{Emitted synchrotron photons}

The photon density in the synchrotron source behind the subshock depends on $\nu_{bb}$.
If $\nu_{bb}>\nu_0$, the source is completely self-absorbed and its spectrum at frequencies 
$\nu<\nu_0$ takes a Raleigh-Jeans form with the temperature $\theta_0=\gamma_0/3$. 
The spectrum has an exponential cutoff at $\nu\sim\nu_0$ and sustains photon number density 
\be
     n_{bb}\approx \frac{8\pi \nu_0^2kT_0}{h c^3}
   =\frac{8\pi \nu_0^2\theta_0}{\lambda c^2}, \qquad \nu_{bb}>\nu_0,
\ee
where $\lambda=h/m_ec$ is the Compton wavelength.

If $\nu_{bb}<\nu_0$, the emitted synchrotron spectrum in the range $\nu_{bb}<\nu<\nu_0$
has the form $I_\nu\propto\nu^{-1/2}$. Its photon number peaks at low frequencies as $\nu^{-1/2}$. 
Therefore, the photon number density inside the source peaks at $\nu_{bb}$ 
and may be written as
\be
   n_{bb}\approx n_0\left(\frac{\nu_0}{\nu_{bb}}\right)^{-1/2},  \qquad \nu_{bb}<\nu_0.
\ee
Here $n_0$ is the number density of photons with frequencies $\nu\sim\nu_0=\gamma_0^2\nu_B$
(see Section~\ref{ssec:pair_loading}),
\be
\label{eq:n0}
     n_0= \frac{u_s(\nu_0)}{h\nu_0}\approx \frac{f g\, u_B}{\gamma_0^2\, h \nu_B\, \xiKN}.
\ee   
The total number density of {\it emitted} photons, $\sim n_{bb}$, is set by $n_0$ and
$\nu_{bb}/\nu_0$. The number density of {\it surviving} synchrotron photons in the RMS, 
$n_s$, is controlled by the positions of $\nu_{abs}$ and $\nu_{ind}$ relative to $\nu_0$.

\subsection{Surviving synchrotron photons}

It is convenient to define
\be
    \nu_1=\max\{\nu_{abs},\nu_{ind}\}.
\ee
Note that $\nu_1>\nu_{bb}$, because $\nu_{ind}>\nu_{bb}$ (Equation~(\ref{eq:nuind})). 
In general, the following four cases are possible.
\\
(1) $\nu_1<\nu_0$. Radiation below $\nu_1$ is suppressed, 
and the number density of surviving photons $\nu_1\simlt\nu\simlt\nu_0$ peaks at 
$\nu\sim\nu_1$,
\be
  n_s\sim n_0\left(\frac{\nu_0}{\nu_1}\right)^{1/2}, \qquad \nu_{abs},\nu_{ind}<\nu_0.
\ee 
(2) $\nu_{ind}<\nu_0<\nu_{abs}$. The source is 
exponentially suppressed by free-free absorption at all frequencies, $\nu\simlt \nu_0$, by the 
factor of $\sim\exp[-\tau_{abs}(\nu)]=\exp[-(\nu_{abs}/\nu)^{2}]$, as the plasma temperature 
is far below the brightness temperature of the synchrotron source.
The suppression is least severe at the upper frequency $\nu_0$, and 
the resulting density of surviving photons is 
\be
   n_s\sim n_0\exp\left(-\frac{\nu_{abs}^2}{\nu_0^2}\right), \qquad \nu_{ind}<\nu_0<\nu_{abs}.
\ee
(3) $\nu_{abs}<\nu_0<\nu_{ind}$. In this case, most photons emitted by the synchrotron source are 
quickly downscattered, so that their frequency moves below $\nu_{abs}$ and they get absorbed. 
The number density of surviving synchrotron photons $n_s$ (with $\nu\sim\nu_0$) is 
estimated from Equation~(\ref{eq:cond}), using the general relation between $n_s$ and 
brightness temperature $\theta_b$,
\be
   n_s(\nu)\approx \frac{u(\nu)}{h\nu}  = \frac{8\pi \nu^2\theta_b}{\lambda c^2}. 
\ee
Equation~(\ref{eq:cond}) gives
\be
\label{eq:nsind}
   n_s\sim \frac{8\pi \nu_0^2\beta_r^2}{\lambda c^2}, \qquad \nu_{abs}<\nu_0<\nu_{ind}.
\ee
(4) $\nu_{abs},\nu_{ind}>\nu_0$. Both free-free absorption and induced downscattering act to suppress
the synchrotron source at all frequencies $\nu\simlt \nu_0$. In this case, one can estimate, 
\be
   n_s\sim \min\left\{ n_0\exp\left(-\frac{\nu_{abs}^2}{\nu_0^2}\right),
   \frac{8\pi \nu_0^2\beta_r^2}{\lambda c^2} \right\}, \quad \nu_{abs},\nu_{ind}>\nu_0.
\ee

\subsection{Photon-to-proton ratio}
\label{ssec:ratio}

The main quantity of interest in this paper is the number of photons produced per 
proton flowing through the shock,
\be
  Z_s=\frac{n_s}{n_p},
\ee
where $n_s$ and $n_p$ are both measured in the downstream rest frame.
Equation~(\ref{eq:n0}) gives
\be
  Z_0=\frac{n_0}{n_p} \approx \frac{f \sigma g}{2\,\xi_{KN}\gamma_0^2}\,\frac{m_pc^2}{h\nu_B}
     \approx \frac{4.4\times 10^6}{B_7(\gamma_0/30)^2}\,\frac{f \sigma g}{\xi_{KN}}.
\label{eq:ns_np_1}
\ee
Note that $Z_0$ depends on both the dimensionless magnetization parameter $\sigma$ 
and the magnetic field strength $B$.
$Z_0$ enters $Z_s$ in a few cases discussed above, depending on the relative
positions of $\nu_{abs},\nu_{ind}$, and $\nu_0$. 

In the case of $\nu_{abs}<\nu_0<\nu_{ind}$, we obtain $n_s/n_p$ from Equation~(\ref{eq:nsind}).
Substituting $\nu_0=(eB/2\pi m_ec)\gamma_0^2$ and using $\sigma=B^2/4\pi n_p m_pc^2$ 
we find
\be
\label{eq:Zind}
  Z_s\sim \Zind= \frac{4}{\pi}\,\alpha_f\,\frac{m_p}{m_e}\,\sigma\,\beta_r^2\gamma_0^4
     \approx 1.4\times 10^7\,\sigma\,\beta_r^2\left(\frac{\gamma_0}{30}\right)^4,
\ee
where $\alpha_f=e^2/\hbar c\approx 1/137$ is the fine structure constant.
Note that $Z_0=\Zind$ when $\nu_{ind}=\nu_0$, and so our result may be written as
\begin{eqnarray}
\label{eq:Zs}
  Z_s\sim\Zind\times \left\{\begin{array}{ll}
                     (\nu_0/\nu_{ind})^{1/2} & \quad \nu_{abs}<\nu_{ind}<\nu_0 \\
                     1   & \quad \nu_{abs}<\nu_0<\nu_{ind}
                                            \end{array}
                                   \right.
\end{eqnarray}
This equation describes the generated photon number when free-free absorption is negligible, 
$\nu_{abs}<\nu_0,\nu_{ind}$. 

The relative positions of $\nu_{abs}$, $\nu_{ind}$, and $\nu_0$ can be found from 
Equations~(\ref{eq:nu_bb}), (\ref{eq:Bbb}), (\ref{eq:nuind}), and (\ref{eq:abs}),
\be
\label{eq:nu_ratio}
    \frac{\nu_{ind}}{\nu_0}\approx  \frac{0.79}{B_7^{1/3}}\, 
     \left(\frac{fg}{\xi_{KN}}\right)^{1/3} \beta_r^{-4/5} \left(\frac{\gamma_0}{30}\right)^{-29/15},
\ee
\be
   \frac{\nu_{ind}}{\nu_{abs}}\approx \frac{85}{B_7^{1/3}}\, 
   \frac{g^{1/3} \sigma^{1/2}\beta_r^{1/5}} {\xi_{KN}^{1/3}f^{1/6}w^{1/2}\theta_{-2}^{3/4}}
    \left(\frac{\gamma_0}{30}\right)^{17/30}.
\ee

\subsection{Relativistic shocks $\beta_r\gamma_r > 1$}

The RMS with $\beta_r\gamma_r >1$ is capable of producing copious pairs through bulk Comptonization of photons, leading to $Z_\pm\simgt 10^2$ (Section~3.3). Then one can see from Equation~(\ref{eq:absZ}) that the synchrotron radiation emitted by the subshock is efficiently 
suppressed by free-free absorption, $\nu_{abs}>\nu_0$, as long as $w$, $\sigma$, and $f$
are somewhat below unity. This suppression occurs because the high $Z_\pm$ pushes the 
characteristic thermal Lorentz factor of the subshock to a low value 
$\gamma_0\approx 14\,f w/Z_{\pm,2}$ (Equation~\ref{eq:gamma_0_Z_pm}),
as the dissipated heat is shared by a large number of $e^\pm$.

A significant number of synchrotron photons can survive in a relativistic RMS if 
its downstream magnetization $\sigma$ is comparable to 
unity,\footnote{The corresponding magnetization of the pre-shock plasma brought by the
   upstream, $\sigma_u$, is related to the downstream $\sigma$ by $\sigma=\xi\sigma_u$, 
   where $\xi$ is the shock compression ratio calculated in B17.}
which also implies a strong subshock $f\simlt 1$.
However, even when $\nu_0>\nu_{abs}$ due to strong magnetization,
$Z_s$ is still limited by induced downscattering (Equation~\ref{eq:Zs}), and the low
$\gamma_0$ in the relativistic RMS implies a modest $\Zind$.

\subsection{Mildly relativistic shocks $\beta_r\gamma_r \simlt 1$}

For RMSs with $\beta_r\gamma_r \simlt 1$ pair creation by energetic photons from bulk 
Comptonization is negligible; instead, pairs are created by the IC photons from the subshock.
A moderate magnetization $\sigma=0.01-0.1$ is sufficient to create a 
strong subshock ($f>0.1$) which emits IC and synchrotron photons. 
The resulting $Z_\pm$ is lower than in relativistic RMSs, and $\gamma_0$ is higher,
$\gamma_0=20-40$ (Section~3.3). Then Equation~(\ref{eq:abs}) shows that $\nu_{abs}$ 
is typically well below $\nu_0$, and then free-free absorption is inefficient. 
The relatively high $\gamma_0\sim 30$ also implies that induced scattering allows 
a high $Z_s\sim\Zind$, even when $\nu_{ind}>\nu_0$. Therefore, mildly relativistic RMSs 
are much more efficient in generating photons compared with relativistic RMSs.

In particular, consider an RMS with $\sigma=0.01-0.1$ propagating in 
a relatively cold medium (a high Mach-number shock). The enthalpy generated by the 
shock is 
\be
  w = \frac{4}{3} (\gamma_r-1).
\ee
Using (\ref{eq:abs}) and $\gamma_0\sim 30$, one can see that $\nu_{abs}<\nu_0$ for 
mildly relativistic shocks $0.2\simlt \beta_r\gamma_r\simlt 1$. For a broad range of 
$B<10^{10}$~G, only induced downscattering can impact the survival of photons, and 
the resulting photon number $Z_s$ is given by Equations~(\ref{eq:Zs}) and (\ref{eq:nu_ratio}), 
which we rewrite as 
\begin{eqnarray}
\label{eq:ZsB}
  Z_s\sim \Zind\times \left\{\begin{array}{ll}
                     (B/B_{ind})^{1/6} & \quad B>B_{ind} \\
                  \;\;\;   1   & \quad B<B_{ind}
                                            \end{array}
                                   \right.
\end{eqnarray}
\be
   B_{ind}\approx 5\times 10^6\,\frac{f\,g}{\xi_{KN}\,\beta_r^{12/5}}
        \left(\frac{\gamma_0}{30}\right)^{-29/5} {\rm ~G}.
\ee
Here $\xi_{KN}\approx 1$ and $g\sim 1$ (Section~3.1); 
$f<1$ increases with $\sigma$ and approaches $\sim 1$ at high magnetizations.

\section{Subphotospheric shocks in GRB jets}

The main parameter of a jet is its total (isotropic equivalent) power,
\be
    L_{tot} = L(1 + w + \sigma), \qquad L=4 \pi r^2 \Gamma^2 n_p m_p c^3,
\ee
where $r$ is radius, and $\Gamma\gg 1$ is the bulk Lorentz factor, and $L$ is the kinetic power 
of the plasma flow in the jet. Here we have neglected the plasma heat compared to its rest 
mass energy. The magnetic field carried by the expanding jet is nearly transverse to the jet velocity. 
Its value measured in the jet rest frame is related to the Poynting flux by 
\be
     B^2=\frac{\sigma L}{c\,r^2\Gamma^2}.
\ee

Instead of the radial coordinate $r$ it will be convenient to use the jet optical depth as an 
independent variable, which is decreasing with radius.
Let $\tau_p$ be the optical depth associated with the ``original'' 
electron-proton plasma flowing in the jet, not counting the created $e^\pm$ pairs. 
The number of original particles is conserved in the expanding jet. At a radius $r$ the
optical depth is given by 
\be
\tau_p = \frac{n_p \sT r}{\Gamma}=\frac{\sT L}{4\pi r \Gamma^3m_pc^3}.
\label{eq:tau_p}
\ee
One can express the magnetic field as a function of $\tau_p$,
\be
   B=\frac{4\pi m_pc^2\Gamma^2\tau_p}{\sT}\left(\frac{c\sigma}{L}\right)^{1/2}
   \approx 4.9\times 10^5\,\frac{\Gamma_2^2\,\sigma^{1/2}}{L_{52}^{1/2}}\,\tau_p.
\ee
The magnetic field in the subphotospheric region $\tau_p\simgt 1$ is not very far from $B_{ind}$,
and may be smaller or larger than $B_{ind}$, depending on the jet parameters. 

As discussed in Section~5.5, synchrotron photons produced by mildly relativistic RMS
do not suffer from free-free absorption. Then the produced photon number per proton 
is described by Equation~(\ref{eq:ZsB}). 
Note that $Z_s$ weakly depends on $B$, and $Z_s\sim Z_\star$ is a good estimate for a
broad range of $B$.
Thus, we conclude that mildly relativistic 
subphotospheric shocks in GRBs generate 
photon number $\sim \Zind\approx 10^6 \sigma_{-1} \beta_r^2(\gamma_0/30)^4$.
This is a large number, which can easily exceed the photon number in the 
unshocked upstream plasma. The shock then feeds itself with photons, which are
bulk Comptonized and mediate the RMS. 

The generated photon number then also sets the average energy of post-shock photons, 
$\epsilon_d$, 
\be
  \epsilon_d\sim \frac{m_p(\gamma_r-1)}{m_e\Zind}\sim 
  \frac{1}{\alpha_f \sigma\gamma_0^4}
   \sim \frac{10^{-3}}{\sigma_{-1}}\left(\frac{\gamma_0}{30}\right)^{-4}.
\ee
The low $\epsilon_d$ implies a low plasma (Compton) temperature $\theta\sim\epsilon_d$.
Nevertheless, free-free absorption remains inefficient, as long as $\theta>10^{-4}$ 
(Equation~\ref{eq:abs}). 

Note also that the characteristic energy of injected synchrotron photons $h\nu_0$ 
(measured in the jet frame) is low,
\be
  \frac{h\nu_0}{m_ec^2}\sim \hbar\,\frac{eB}{m_e^2c^3}\gamma_0^2 
     \approx 10^{-5}\, \frac{\Gamma_2^2\,\sigma^{1/2}}{L_{k,52}^{1/2}}\,
     \left(\frac{\gamma_0}{30}\right)^2 \tau_p.
\ee

\section{Summary}
\label{sec:discussion}

Our main results may be summarized as follows:

(1) We have shown that RMS propagating in a moderately magnetized plasma is capable of 
producing copious synchrotron photons, because there is a collisionless subshock in the magnetized RMS.The subshock heats pairs to a relativistic temperature, 
and the fast-cooling pairs emit a fraction of their energy as synchrotron photons.

(2) The synchrotron emission occurs at rather low frequencies and is therefore prone to self-absorption. Another threat to synchrotron photons is free-free absorption, which is helped by induced downscattering in the plasma flowing through the RMS. 
We find that these processes suppress
the synchrotron source in relativistic RMSs with $\beta_r\gamma_r \simgt 1$. 
This is because such shocks generate a large amount of $e^\pm$ pairs which share 
the subshock energy, with a reduced energy per particle, leading to a particularly low
characteristic frequency of the synchrotron source $\nu_0$.

(3) Mildly relativistic RMSs ($\beta_r\gamma_r \simlt 1$) efficiently produce photons. The resulting photon number generated per proton flowing through the shock is 
$Z\sim 10^6\sigma_{-1}\beta_r^2 (\gamma_0/30)^4$.
Here $\gamma_0$ is the thermal Lorentz factor of shock-heated electrons, which 
is self-regulated by pair creation to $\gamma_0\sim 30$.

(4) The generated photons can easily dominate over pre-shock photons in the upstream plasma.
Then the RMS is essentially feeding itself with photons, which are upscattered through bulk
Comptonization and mediate the shock.

One implication of our results is a preferred range of subphotospheric 
magnetization in GRBs, $\sigma=0.01$-$0.1$, as it gives the photon number 
consistent with observations, $Z_{obs}\sim 10^5-10^6$ (e.g. Beloborodov 2013). 
Higher $\sigma>0.1$ would 
lead to overproduction of photons, reducing the spectral peak energy well below 
$E_{peak}\sim 1$~MeV typical for GRBs. Our preferred range of magnetization is also 
consistent with $\sigma$ obtained from fitting GRB spectra by photospheric radiative 
transfer models (Vurm \& Beloborodov 2016).

The emission of photons  in subphotospheric shocks will affect their radiation spectra.
With the increased number of soft photons the efficiency of photon energy gain via bulk Comptonization is lowered 
(the RMS transition somewhat spreads out), 
and the power-law slope of the nonthermal radiation inside the shock somewhat softens.
A small bump at the synchrotron injection energy $\sim h\nu_0$ may appear at the low-energy end
of the spectrum.

\acknowledgments
CL acknowledges the Swedish Research Council for financial support. AMB is supported by 
NASA grant NNX15AE26G and a grant from the Simons Foundation (\#446228, Andrei Beloborodov).

\bibliographystyle{apj}
\bibliography{refgrb}


\appendix

\section{Bremsstrahlung and double Compton photon production}
\label{app:BrDC}

The photon production rates per unit volume for bremsstrahlung and double Compton emission
can be written as \citep{Sve:1984}

\be
\frac{\mathrm{d}\dot{n}_\gamma}{\mathrm{d}\ln\epsilon} = \alpha_f r_e^2 c n_1 n_2 F(\epsilon, \theta),
\label{eq:dndlneps}
\ee

\noindent where $\alpha_f$ is the fine structure constant, $r_e$ is the classical electron radius, $n_1$ and $n_2$ are the number densities of the particles involved, and $F(\epsilon, \theta)$ is a function that describes the energy and temperature ($\theta \equiv kT/m_e c^2$) dependence of the radiative process. Function $F$ depends on $\epsilon$ logarithmically for bremsstrahlung and is independent of $\epsilon$ for double Compton.

The characteristic timescale for the RMS is the photon mean free path time to Thomson scattering, $t_{sc}=(Z_\pm n_p \sigma_T c)^{-1}$. We define a dimensionless photon production rate as

\be
\xi \equiv \frac{t_{sc}}{n_\gamma}\, \dot{n}_\gamma
= \frac{3\alpha_f}{8\pi} G F \frac{n_1 n_2}{Z_\pm n_p n_\gamma},
\ee

\noindent where 
$n_\gamma$ is the density of photons advected from the upstream, and $\xi > 1$ would indicate significant photon production over a scattering time. The weak (or non-existent) energy dependence of $F$ makes it convenient to write $\dot{n}_\gamma = G (\mathrm{d}\dot{n}_\gamma/\mathrm{d}\ln\epsilon)$. Here $G\sim 10$ results from
the integration of the photon production rate over several decades in energy where $\mathrm{d}\dot{n}_\gamma/\mathrm{d}\ln\epsilon$ is essentially constant. (The lower limit of the integral is set by the requirement that the photon can double its energy by scatterings before being absorbed.) The relevant number densities are $n_1 n_2 = n_\pm n_\gamma = Z_\pm n_p n_\gamma$ for double Compton, $n_1 n_2 = n_+ n_- = (Z_\pm^2 - 1) n_p^2 / 4$ for electron-positron bremsstrahlung, and $n_1 n_2 = n_\pm n_p = Z_\pm n_p^2$ for pair-proton bremsstrahlung. We then find

\be
\xi_{DC} = \frac{3\alpha_f}{8\pi} G F_{DC}, \qquad
\xi_\pm = \frac{3\alpha_f}{32\pi} \frac{Z_\pm^2 - 1}{Z_\pm} \frac{n_p}{n_\gamma} G F_\pm, \qquad
\xi_{\pm p} = \frac{3\alpha_f}{8\pi} \frac{n_p}{n_\gamma} G F_{\pm p}.
\ee

\noindent Complete expressions for the functions $F_{DC}$, $F_\pm$ and $F_{\pm p}$ are given by \citet{Sve:1984}; the low temperature limits ($\theta \ll 1$) are

\be
F_{DC} \approx \frac{128}{3}\frac{\theta^2}{1 + 13.91 \theta}, \qquad
F_\pm \approx \frac{64}{3} \ln \left(\frac{4\theta}{\epsilon}\right) \frac{1}{(\pi\theta)^{1/2}}, \qquad
F_{\pm p} \approx \frac{32}{3 \sqrt{2}} \ln \left(\frac{4\theta}{x} \right) \frac{1}{(\pi \theta)^{1/2}},
\ee

\noindent where $F_{DC}$ was computed assuming that the photons have a Wien spectrum and that the radiation and electron temperatures are equal. The function $F$ (and therefore also $G$) depends on the electron temperature and the photon spectrum (c.f. \citealt{Sve:1984}, Equation~(A7)). 
It is most sensitive to the electron temperature and the average photon energy, and
not to the exact spectral shape.

\begin{figure}
\centering
\includegraphics[width=.5\linewidth]{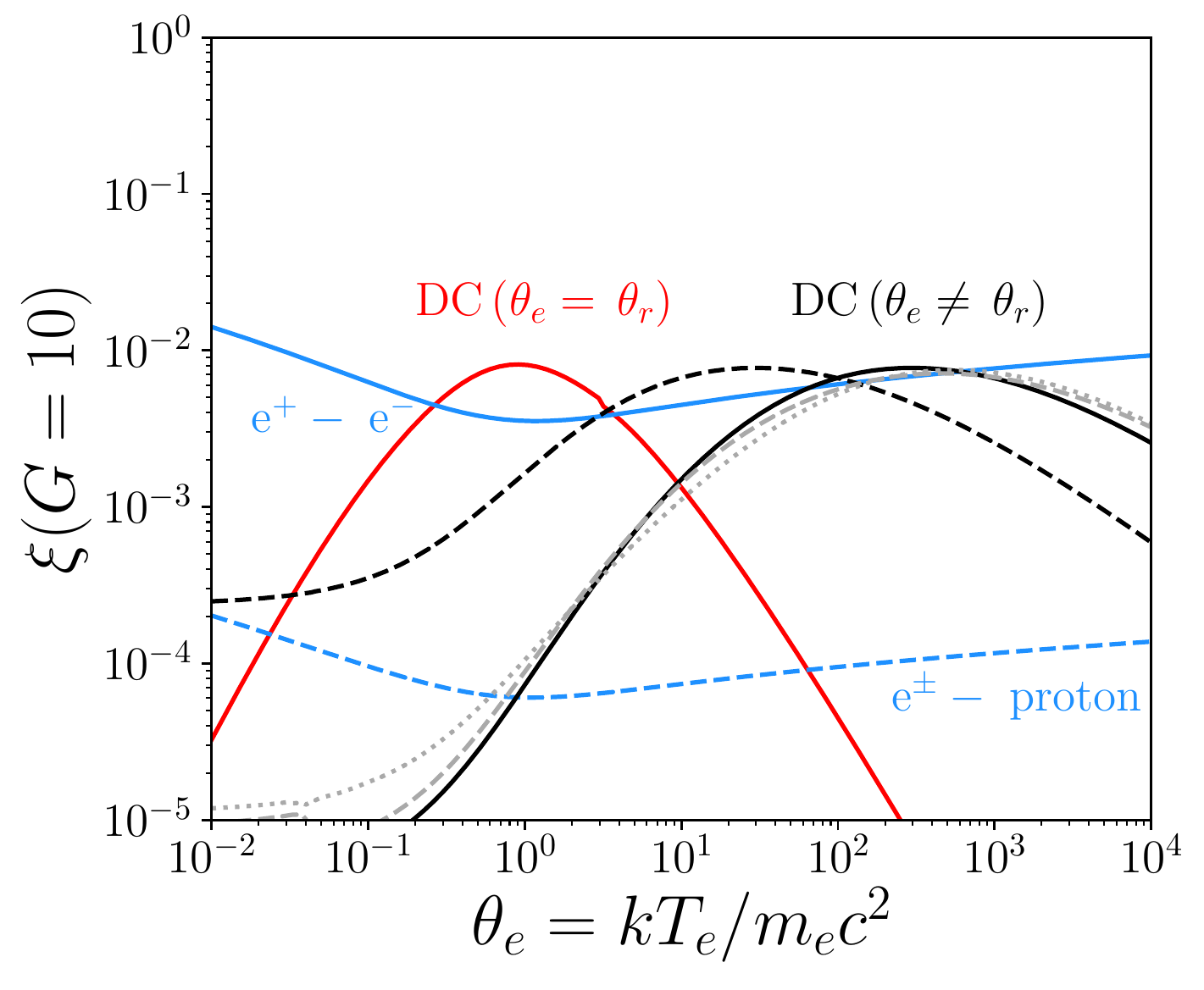}
\caption{Normalized emission rates for double Compton and bremsstrahlung as a function of the electron temperature. The blue lines show bremsstrahlung emission, while the red line shows double Compton emission from a Wien spectrum with the radiation temperature $T_r$ equal to the electron temperature $T_e$. The black lines are numerical integrations (c.f. \citealt{Sve:1984}, Equation~(A7)) and assume Wien spectra with a fixed average photon energy of $\bar{\epsilon} = 10^{-2}$ (solid line) and $\bar{\epsilon} = 10^{-1}$ (dashed line); in these cases $T_r \neq T_e$. The gray lines (also numerical integrations) assume power law photon spectra, either flat in $\nu F_\nu$ (dashed line) or flat in $F_\nu$ (dotted line), ranging about two decades in photon energy and with average photon energies of $\bar{\epsilon} = 10^{-2}$. The black and gray lines are applicable to subshocks, where electrons cool in an ``external'' radiation field of fixed average photon energy. (A value of $G = 10$ was used for this plot, corresponding to emission being absorbed at about five orders of magnitude below the emission peak).}
\label{fig:DC_Br}
\end{figure}

Figure~\ref{fig:DC_Br} shows the normalized emission rates $\xi_{DC}$, $\xi_\pm$ and $\xi_{\pm p}$ as functions of the electron temperature. A large pair multiplicity of $Z_\pm = 10^2$ and a small photon to baryon ratio of $n_\gamma/n_p = 10^4$ was used in order to provide favorable conditions for bremsstrahlung photon production. The double Compton rate is shown for various conditions, relevant for the RMS or its subshock. Note that $F_{DC}$ is always less than unity, and $\xi_{DC}$ is suppressed by the fine structure constant and is therefore always less than $10^{-2}$. It is clear that $\xi \ll 1$ for any temperature relevant for GRB jets.

This conclusion may not be valid for non-relativistic shocks $\beta_r\ll 2$ where the time given to photon production (the plasma crossing time of the RMS) is increased to $\sim t_{sc}/\beta_r^2$.
Photon production could then be effective, however such weak shocks barely generate any entropy, and thus there is essentially no need for photon production. 

As can be seen in Figure~\ref{fig:DC_Br}, high temperatures do not help the photon production rate much. Therefore the hot plasma immediately behind the subshock in an RMS is also incapable of generating photons via bremsstrahlung or double Compton scatterings.

\end{document}